\documentclass[%
 aip,
 amsmath,amssymb,
 reprint,%
]{revtex4-2}
\usepackage{gensymb}
\usepackage{tabularx}
\usepackage{caption}
\usepackage{booktabs}
\usepackage{subfig}
\usepackage{graphicx}
\usepackage{dcolumn}
\usepackage{bm}
\usepackage{xcolor}
\usepackage[utf8]{inputenc}
\usepackage[T1]{fontenc}
\usepackage{mathptmx}
\usepackage{etoolbox}
\usepackage{float}
\usepackage{enumitem}
\begin{document}

\preprint{AIP/123-QED}

\title[Influence of thermal effects on the optomechanical coupling rate in acousto-optic cavities]{Influence of thermal effects on the optomechanical coupling rate in acousto-optic cavities}
\author{Raúl Ortiz}%
 \affiliation{Nanophotonics Technology Center, Universitat Polit\`ecnica de Val\`encia, Camino de Vera s/n, 46022 Valencia, Spain}
\author{Laura Mercad\'e}
\affiliation{Nanophotonics Technology Center, Universitat Polit\`ecnica de Val\`encia, Camino de Vera s/n, 46022 Valencia, Spain}
\affiliation{MIND-IN2UB, Departament d’Enginyeria Electrònica i Biomèdica, Facultat de Física, Universitat de Barcelona, Martí i Franquès 1, Barcelona 08028, Spain}
\author{Alberto Grau}
\affiliation{Nanophotonics Technology Center, Universitat Polit\`ecnica de Val\`encia, Camino de Vera s/n, 46022 Valencia, Spain}
\author{Daniel Navarro-Urrios}
\affiliation{MIND-IN2UB, Departament d’Enginyeria Electrònica i Biomèdica, Facultat de Física, Universitat de Barcelona, Martí i Franquès 1, Barcelona 08028, Spain}
\author{Alejandro Mart\'inez}
\affiliation{Nanophotonics Technology Center, Universitat Polit\`ecnica de Val\`encia, Camino de Vera s/n, 46022 Valencia, Spain}
\email{amartinez@ntc.upv.es}

\date{\today}

\begin{abstract}
Optomechanical (OM) cavities simultaneously localize photons and phonons, thus enhancing their mutual interaction through radiation-pressure force. This acousto-optic interaction can be quantified by means of the optical frequency shift per mechanical displacement $G$. The aforesaid frequency shift can also be related to the vacuum OM coupling rate, $g_{0}$, where only photoelastic (PE) and moving boundaries (MB) effects are commonly taken into account. However, the thermo-optic (TO) and thermal expansion (ThE) effects may also play a role since the material forming the OM cavity could be heated by the presence of photons, which should naturally affect the mechanical properties of the cavity. In this work, we introduce a new theoretical approach to determine how thermal effects change the canonical OM coupling rate. To test the model, a complete set of optical-thermal-mechanical simulations has been performed in two OM crystal cavities fabricated from two different materials: silicon and diamond. Our results lead us to conclude that there is a non-negligible thermal correction that is always present as a negative shift to the OM coupling rate that should be considered in order to predict more accurately the strength of the OM interaction.
\end{abstract}

\maketitle


\section{Introduction}
Cavity optomechanics is a growing scientific and technological field that encompasses assorted micro- and nano-devices that enable the interaction between optical and mechanical modes giving rise to various striking phenomena \cite{Chan2011,Verhagen2012,AKM14_RMP,Pennec2014,NAV17_NCOMMS,Verhagen22,https://doi.org/10.1002/lpor.202100175,doi:10.1126/science.1195596}. Optomechanical (OM) interaction in such devices is mediated by the radiation pressure force that enables the exchange of energy and momentum between photons and phonons. Due to the retarded nature of radiation pressure force, a dynamical backaction between the optical and mechanical modes takes place \cite{Kippenberg2008}, resulting in the optical spring effect \cite{Vogel2003} as well as the cooling (heating) of the OM cavity when the driving laser is red(blue)-detuned with respect to the optical resonance \cite{Chan2011,Liu2013,MER20_NP}. To quantify this interaction, the vacuum OM coupling rate, $g_0$, is introduced, which accounts for the coupling between a single photon and a single phonon \cite{AKM14_RMP}, thus resulting in the most commonly used value for comparing the strength of the acousto-optic interaction in OM cavities.  \\
\indent Accurate calculations of $g_0$ are necessary to predict all the aforementioned effects in simulations as well as to optimize OM cavities. So far, only the photoelastic (PE) and moving boundaries (MB) effects have been considered to contribute to the calculation of $g_0$ i.e., only purely acousto-optic effects have been taken into account \cite{Pennec2014,MBPE,Djafari}. To the best of our knowledge, how thermal effects contribute to the canonical OM coupling rate has not been considered before. In fact, such thermal effects may eventually have an impact on the value of $g_0$ as previous works have reported experimental variations of this parameter with temperature \cite{Jiang:19,FaveroPTH}. 


\indent
In that regard, a model that describes photothermal forces through a mathematical treatment built upon thermal modal analysis and perturbation theory was recently proposed and checked experimentally in a GaAs microdisk OM cavity\cite{Primo2021}. In that model, the photothermal and the OM backactions were simultaneously introduced in the OM structure, leading to dynamical equations mediated by thermal Langevin, radiation pressure, and photothermal forces. In addition, a thermally-induced frequency shift dependence was addressed theoretically by using perturbation theory and described with a new opto-thermal pull parameter $G^{\theta} = \frac{\partial \omega}{\partial T}$.  However, the impact of the temperature on the original OM pull parameter, $G^{OM}=\frac{\partial \omega}{\partial \alpha}$, was not considered, although it would induce deformations and frequency shifts in the cavity. Thermal effects, together with the acousto-optic PE and MB effects must be then considered in order to obtain accurate calculations of the $g_0$ parameter.\\
\indent
In this work, we develop a theory that considers the thermo-optic (TO) and thermal expansion (ThE) effects, which account for the optical resonance shift due to temperature absorption and thermal expansion in the material, respectively. These effects are incorporated as corrections to the calculation of the OM coupling rate $g$ and vacuum OM coupling rate $g_0$ in OM crystal cavities. Theoretical work is combined with numerical simulations in silicon and diamond OM crystal cavities using \textit{Comsol Multiphysics}. Numerical results confirm the theoretical predictions and highlight the need to consider the new $g^{Th}_0$ term in particular in materials showing large thermo-optical coefficients. 

\section{Theoretical approach}

\begin{figure}
    \centering
    \includegraphics[width=0.7\columnwidth]{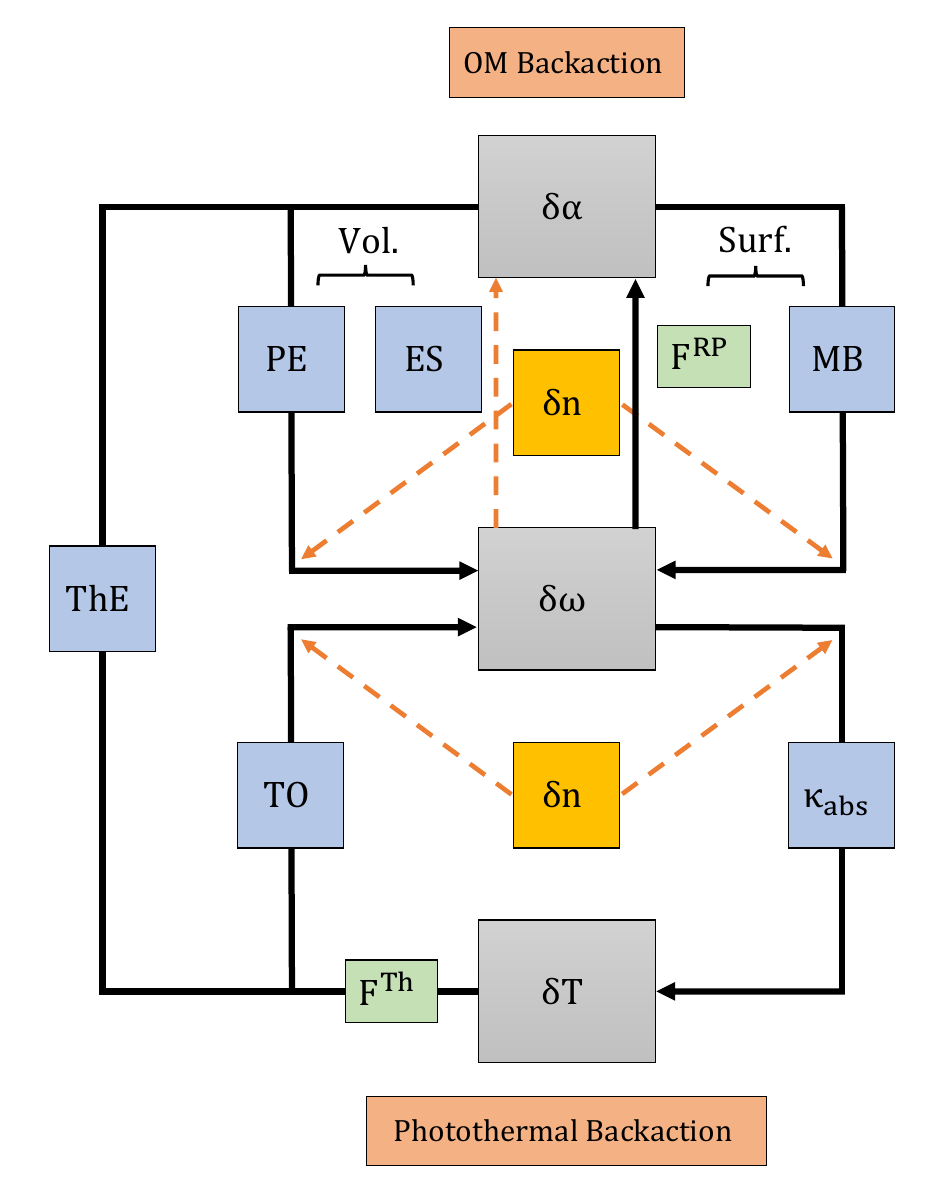}
    \caption{OM and photothermal dynamical backaction loops. OM backaction is mediated by radiation pressure force ($F^{RP}$) and electrostriction (ES) from optical to mechanical modes and by photoelastic (PE) and moving boundaries (MB) effects from mechanical to optical modes. The whole process is coupled causing fluctuations in the optical and mechanical resonances. Photothermal backaction is mediated by photothermal forces which include both thermo-optic (TO) and thermal expansion (ThE) effects. The first one closes the photothermal backaction whereas the second connects to the OM backaction.}
    \label{fig:loop}
\end{figure}

Figure \ref{fig:loop} sketches the dynamical backactions loops arising from the OM and photothermal effects taking place in an OM cavity, and how they are connected by the thermal expansion (ThE) effect. In the OM dynamical backaction picture, mechanical modes and deformations $\alpha$ are excited by the optical energy in the cavity by means of the radiation pressure force and electrostriction (ES) effects. Photons can exchange energy and momentum with the material creating coherent phonons at frequencies of the mechanical modes of the cavity. Once a mechanical mode is excited, it exchanges energy back to the optical mode inducing fluctuations in the optical resonance by two different phenomena: photoelastic (PE) and moving boundaries (MB) effects \cite{Pennec2014}. In the former, mechanical vibrations and deformations ($\delta \alpha$) change the stress in the material thus modifying the refractive index and therefore producing fluctuations in the optical resonance frequency $\delta\omega$. In addition, excitation of the mechanical modes changes dynamically the position of the boundaries of the cavity, thus provoking a variation of the refractive index and, consequently, changing the optical mode (MB effect). This closes the OM dynamical backaction loop. It is worth noting that (ES) and (PE) are volumetric acousto-optic effects, whereas $F^{RP}$ and (MB) are surface effects \cite{Yan-Nano}. 

Conversely, photothermal backaction follows a predominantly unidirectional path, characterized by a loop that progresses solely in one direction. Refractive index variations produced by tiny optical fluctuations produce an optical absorption pattern that depends on the optical mode profile. It makes the material absorb the optical power and increases its temperature creating a gradient profile. This gradient produces photothermal forces that can be described by two phenomena: the thermo-optic (TO) and the thermal expansion (ThE) effects. TO refers to the phenomenon where changes in temperature produce variations in the refractive index of a material causing a red-shift in the optical resonance. This effect has been widely explored in OM cavities and resonators \cite{Lydiate:17}. On the other hand, ThE refers to the tendency of a material to expand or contract in response to changes in temperature \cite{Boyd}. The former closes the photothermal backaction loop, whereas the latter connects with the OM dynamical backaction contributing to both PE and MB effects.  \\
\indent

Accounting for the thermal effects in an OM cavity, its optical resonant frequency can be generally parameterized as: 

\begin{equation}
\omega=\omega(\alpha, T)
\label{eq:principle}
\end{equation}

where $\alpha$ is the overall scaling parameter of the generalized coordinate of the displacement field and $T=T(\vec r)$ is the temperature profile. Because of the interaction between the optical and the mechanical mode, the optical resonance depends on the mechanical resonance as follows:

\begin{equation}
    \omega(\alpha, T)\approx\omega\left(\alpha_0,T_0\right)+\alpha\left(\frac{\partial \omega}{\partial \alpha}\right)_T+\left(T-T_0\right)\left(\frac{\partial \omega}{\partial T}\right)_\alpha
\end{equation}

Here, $\alpha_0$ is the initial maximum amplitude of the material due to mechanical amplitude or deformation of the cavity and $T_0$ temperature in the cavity.

Compared with the typical mechanical contribution to the optical frequency \cite{AKM14_RMP}, we have added a thermal contribution that changed the optical resonance. Thus, we can take the full derivative of the cavity resonance to obtain the OM pull parameter as follows:

\begin{equation}
\frac{d \omega(\alpha, T)}{d \alpha}=\left(\frac{\partial \omega}{\partial \alpha}\right)_T+\left(\frac{\partial \omega}{\partial  T}\right)_\alpha \frac{d T}{d \alpha}
\label{Derivative_eq}
\end{equation}

We can identify the first term on the right side, $\left(\frac{\partial \omega}{\partial \alpha}\right)_T$, as the contribution of the PE and MB effects to the OM coupling, $\left(\frac{\partial \omega}{\partial \alpha}\right)_T = G^{PE} + G^{MB}$. Likewise, we can consider the second term, $\left(\frac{\partial \omega}{\partial  T}\right)_\alpha \frac{d T}{d \alpha} = G^{Th}$, as the thermal contribution to the overall OM coupling rate, $G = \frac{d \omega(\alpha, T)}{d \alpha} = G^{PE} + G^{MB} + G^{Th}$. Specifically, $\left(\frac{\partial \omega}{\partial  T}\right)_\alpha$ is the thermo-optic (TO) effect, whilst $\frac{d T}{d \alpha}$ is the thermal expansion (ThE) term.\\

Therefore, the new thermal term can be interpreted as a correction of the canonical OM pull parameter. It is indeed the contribution that accounts for how thermal deformations and optical frequency shifts due to such thermal effects alter the OM pull parameter of the OM cavity. \\

\indent 
The thermal term of the vacuum OM coupling rate $g_{0}^{Th}$ is related to $G^{Th}$  as:
\begin{equation}
\frac{g_{0}^{Th}}{2 \pi}=\frac{X_{zpf}}{2 \pi}\left(\frac{\partial \omega}{\partial T}\right)_\alpha \frac{d T}{d \alpha}
\label{eq:g}
\end{equation}
where $X_{zpf}$ is the zero-point fluctuation displacement of the cavity mechanical mode. We must highlight that the previous equations have considered the interaction between one photon and one phonon. However, if we increase the power so that the number of intracavity photons is $\bar{n}_{cav}>>1$, the parameter that we obtain is $g^{Th}$. So far, the total vacuum contribution to the OM coupling rate $g_0$ was obtained by normalizing to the total number of photons when $\bar{n}_{cav}>>1$ \cite{AKM14_RMP}. Then, as we consider the thermal contribution as an additive correction, we also have to make this normalization so $g^{Th}_0$ is recalculated as follows:
\begin{equation}
g^{Th}=g_0^{Th} \sqrt{\bar{n}_{cav}}
\label{g0g}
\end{equation}
where $\bar{n}_{cav}$ is related to the intracavity power $P$, the detuning between the laser and the optical cavity frequency $\Delta$, the optical linewidth $\kappa$ and extrinsic optical losses $\kappa_{ex}$\cite{AKM14_RMP}:
\begin{equation}
\overline{n}_{cav}=\frac{\kappa_{ex}}{\Delta^2+(\kappa / 2)^2} \frac{P}{\hbar \omega_{L}}\approx\frac{4 \beta Q P}{\omega^2 \hbar}
\label{fig:ncav}
\end{equation}
Here, we have considered $\kappa_{ex}$ as a fraction of the value of $\kappa$, $\kappa_{ex}=\beta\kappa$. In addition, $\kappa$ is related to the cavity optical quality factor $Q$ as $\kappa=\omega/Q$. For simplicity, we have assumed $\Delta=0$ ($\omega_L= \omega$) which means that we are always measuring the shift in $g^{Th}_0$ exactly at the cavity resonance, although it changes with the power due to the TO effect.\\
From all the previous assumptions, we can write the expression of $g^{Th}_0$ as:

\begin{equation}
\left(\frac{g_0^{Th}}{2 \pi}\right)_{\bar{n}_{cav}>1} = \frac{X_{zpf}}{2 \pi}\left(\frac{\partial \omega}{\partial T}\right)_\alpha \frac{dT}{d \alpha} \omega \sqrt{\frac{\hbar}{4 \beta Q}} \frac{1}{\sqrt{P}}
\label{eq:g0}
\end{equation}

\section{Silicon and diamond OM cavities}
\begin{figure}
    \centering
    \includegraphics[width=\columnwidth]{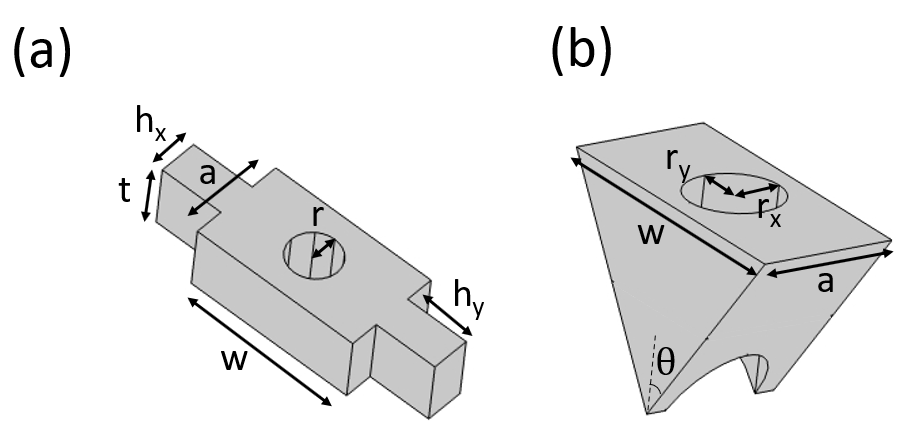}
    \caption{Defect unit cells of the silicon (a) and diamond (b) OM crystal cavities studied in this work. The silicon unit cell has a lattice period $a=325\;nm$, width $w=570\;nm$, thickness $t=220\;nm$, radius $r=84.5\;nm$ and lateral corrugation dimensions $h_x=130\;nm$ and $h_y=215\;nm$. On the other hand, for the diamond cavity, we have $a=464\;nm$, $w=929\;nm$, $r_x=163.5\;nm$, $r_y=144.5\;nm$ and $\theta=35\degree$.}
    \label{Unit_Cells}
\end{figure}

To test our model via numerical simulations, we have chosen released OM crystal cavities made of two different materials: silicon and diamond. It is worth noting that although we do not perform an experimental study in this paper, both cavities have been fabricated and experimentally tested\cite{MER20_NP,Burek:16}. In principle, one may expect a larger thermal response in the silicon cavity since both TO and ThE effects are larger in silicon than in diamond \cite{1997115}. The unit cells employed to build both kinds of OM cavities are depicted in Fig. \ref{Unit_Cells}. For silicon,  the unit cell (Fig. \ref{Unit_Cells}(a)) consists of a circular hole and two lateral corrugations and it is designed to have a TE photonic bandgap and a full phononic bandgap \cite{Jordi_Nature,MER20_NP,Oudich_PRB,Engineering}. On the other hand, the diamond unit cell\cite{Burek:16} (Fig. \ref{Unit_Cells}(b)) has a triangular cross-section parametrized by the etch angle $\theta$ and it has elliptic instead of circular holes. The values of the unit cell parameters in the central region of the cavities employed in the simulations are given in the caption of Fig. \ref{Unit_Cells}.  \\

\begin{figure}
    \centering
    \includegraphics[width=\columnwidth]{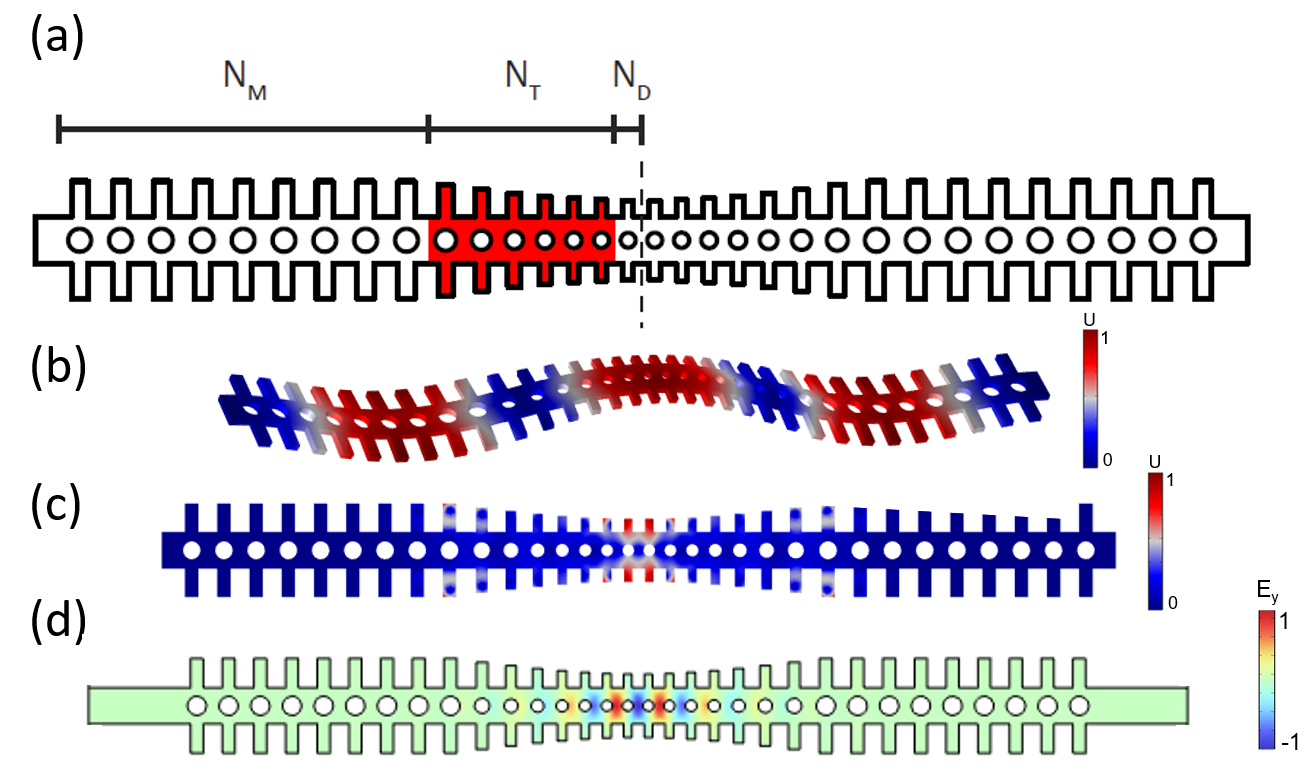}
    \caption{Silicon OM crystal cavity with optical and mechanical modes. (a) Top-view of the cavity where the geometry is divided into three different parts; the defect cells (where the optical and mechanical modes exist), the mirror cells (where the desired modes are reflected), and the transition cells (taper between defect and mirror cells to smooth the geometry and avoid radiative losses). A quadratic tapering along six transition cells is applied to the parameters of the cells to finally reach the mirror values: ($a_m=500\;nm$, $r_m=150\;nm$, $h_{xm}=200\;nm$, $h_{ym}=465\;nm$); Normalized mechanical displacement of the (b) flexural MHz-scale mechanical mode ($\Omega = 15.36\;MHz$) and the (c) defect GHz-scale $\Omega/2\pi=4.01\; GHz$} mechanical mode; (d) Electric field patten of the optical mode of the cavity at $\lambda = 1570\;nm$, which can be coupled to both mechanical modes by OM interaction.
    \label{fig:allmodes}
\end{figure}

\indent
Figure \ref{fig:allmodes}(a) shows a sketch of the silicon OM cavity, which in essence consists of a one-dimensional photonic and phononic crystal nanobeam cavity in which a TE optical mode ($\lambda \approx 1570\;nm$) and a mechanical mode ($\Omega  \approx 4\; GHz$) are well confined in the central region of the nanobeam  (see Fig. \ref{fig:allmodes}(c) and (d)). This colocalization results in a large OM coupling rate\cite{Engineering}. In addition, a mechanical flexural mode oscillating at $\Omega \approx 15\;MHz$ Fig.\ref{fig:allmodes}(b) can be coupled with the same optical mode \cite{Navarro15}.

\begin{figure}
    \centering
    \includegraphics[width=\columnwidth]{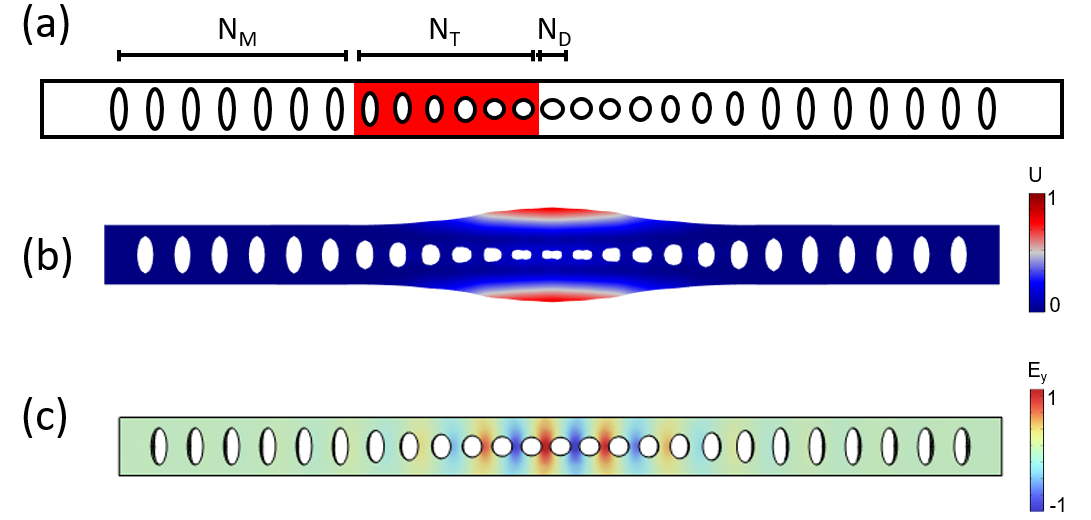}
    \caption{Diamond OM crystal cavity with optical and mechanical modes. (a) Sketch of the diamond OM cavity, where the geometry is divided into three different parts; the defect cells (where the optical and mechanical modes exist), the mirror cells (where the desired modes are reflected), and the transition cells (taper between defect and mirror cells to smooth the geometry and avoid radiative losses). A quadratic tapering along six transition cells is applied to the parameters of the cells to finally reach the mirror values: ($a_m=580\;nm$, $r_{xm}=125\;nm$, $r_{ym}=295\;nm$); (b) Normalized displacement of the mechanical mode oscillating at $\Omega/2\pi=5.95\;GHz$; (c) Electric field of the optical mode of the cavity at $\lambda = 1470\;nm$.}
    \label{fig:Montaje_2}
\end{figure}

On the other hand, we considered an OM crystal cavity constructed from diamond [110] that was previously addressed in \cite{Burek:16}. The structure of this cavity is similar to the previous silicon one: there is a defect cell and two lateral regions composed of an array of mirror cells with a photonic and phononic bandgap (Fig.\ref{fig:Montaje_2}(a)). There are also transition cells that prevent the leakage of photons and phonons out of the cavity. A main difference is that in the diamond cavity the circular holes are replaced with ellipses and there are no lateral corrugations. In this case, the cavity displays a breathing-like mechanical mode at $\Omega \approx 6\;GHz$ (\ref{fig:Montaje_2}(b)) excited by an optical mode at $\lambda \approx 1470\;nm$ (\ref{fig:Montaje_2}(c)). 
The mechanical properties of both cavities can be found in \cite{Burek:16,Adachi}.

\section{Thermal simulations}
\begin{figure*}[ht]
    \centering
    \includegraphics[width=0.8\textwidth]{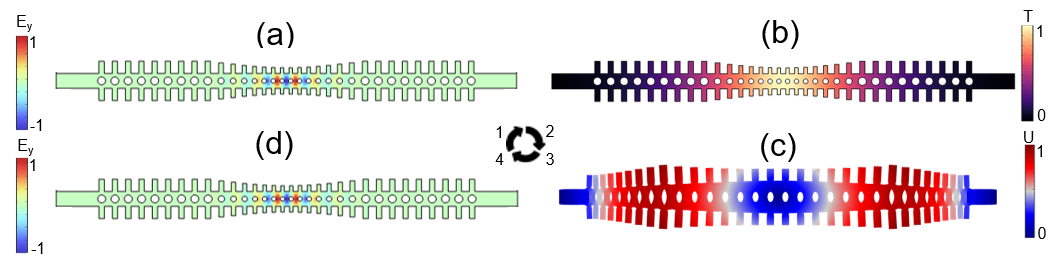}
    \caption{Opto-thermo-mechanical loop in the silicon OM crystal cavity. (a) Optical eigenfrequency profile at $\lambda=1570\;nm$; (b) Thermal field provoked by using (a) as a heating source; (c) Mechanical deformations induced by the thermal profile in (b); (d) Re-calculation of the optical frequency mode including the mechanical deformations.}
    \label{fig:Loop_TOM}
\end{figure*}
\begin{figure*}[ht]
    \centering
    \includegraphics[width=0.8\textwidth]{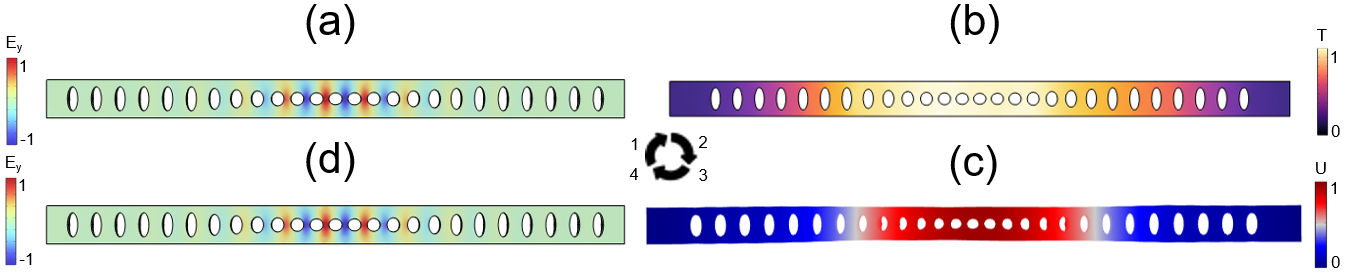}
    \caption{Opto-thermo-mechanical loop in the diamond OM crystal cavity. (a) Optical eigenfrequency profile at $\lambda=1470\;nm$; (b) Thermal field provoked by using (a) as a heating source; (c) Mechanical deformations induced by the thermal profile in (b); (d) Re-calculation of the optical frequency mode including the mechanical deformations..}
    \label{fig:Loop_TOM_Diamond}
\end{figure*}

All the simulations were performed using the numerical tool \textit{Comsol Multiphysics}. In order to model the thermal contribution, we split the study into thermo-optic and thermal expansion sub-studies. Both thermal derivatives in Eq.\ref{Derivative_eq} were obtained by simulating the frequency shift and the thermal expansion of the cavity for several input powers. Each simulation corresponds to a point in both graphs $\omega-T$ and $\alpha-T$. Hence, by fitting the best curve of the sequence of points, one can obtain the derivative of that curve that coincides with the TO $\left(\frac{\partial \omega}{\partial  T}\right)_\alpha$ and ThE $\frac{d T}{d \alpha}$ terms.\\
\indent
If only PE and MB effects are considered, an eigenmode analysis allows to obtain the desired optical and mechanical modes, and then, by using the coupling equations shown in\cite{Chan_Th}, one can calculate the PE and MB contributions to $g_0$. 
Now, with the aim of including the thermal part, we have to add the corrective thermal term to the PE and MB calculations. To do so, the steps in Fig.\ref{fig:Loop_TOM} and Fig.\ref{fig:Loop_TOM_Diamond} must be followed. The first step consists of performing an eigenmode analysis to obtain the optical mode frequency and its profile. The obtained modes (see Fig.\ref{fig:Loop_TOM}(a) and Fig.\ref{fig:Loop_TOM_Diamond}(a) for the silicon and the diamond OM cavity, respectively) seem to be well confined in the defect region of the cavity. We then use the optical mode profile as a heat source (see Fig.\ref{fig:Loop_TOM}(b) and Fig.\ref{fig:Loop_TOM_Diamond}(b)) by assigning a distributed power to it, following the next equation: 

\begin{equation}
Q=Q_0\frac{Q_h}{\int Q_h \;dV}
\label{eq:int}
\end{equation}

Here, $Q$ is the average power weighted by the power density distribution in the domain, measured in $W/m^3$, and $Q_0$ is the input power (measured in $W$), which is distributed into the cavity considering $Q_h$ (the power dissipation density measured in $W/m^3$).

The temperature profile is then modeled by the heat transfer equations as 
\cite{Heat_Transfer}:
\begin{equation}
\begin{aligned}
&\rho C_p u \cdot \nabla T+\nabla \cdot q=Q \\
&q=-k \nabla T
\end{aligned}
\label{Eq:thermal}       
\end{equation}

Here, $\rho$ is the density of the material, $C_p$ is the solid heat capacity at constant pressure, $k$  is the solid thermal conductivity, $u$ is the velocity field and $Q$  is the heat source which coincides with the $Q$ calculated in Eq.\ref{eq:int}. We highlight that we have considered the temperature dependence of $\rho$, $k$, $C_p$ as well as the refractive index $n$ in the numerical simulations to make the heat study accurate \cite{McCaulley1994,Shanks1963}. \\

\indent 
Now, taking a stationary thermal study in which Eq.\ref{Eq:thermal} is solved, we perform a mechanical mode analysis to apply a thermal deformation to the cavity (see Fig.\ref{fig:Loop_TOM}(c) and Fig.\ref{fig:Loop_TOM_Diamond}(c)). Due to the geometry differences, the largest deformation in the silicon OM cavity is concentrated at the transition and mirror cells whereas for the diamond cavity, the deformation is localized at the center. Then, the "moving mesh" interface in \textit{Comsol} is used to store the position of each deformed node, i.e. to obtain the thermally-deformed cavity. Finally, the optical mode is re-calculated in the deformed cavity by performing an eigenmode study (Fig.\ref{fig:Loop_TOM}(d) and Fig.\ref{fig:Loop_TOM_Diamond}(d)), allowing us to parametrize frequency shifts due to thermal effects and completing the loop.\\
\indent
To summarize, these are the paths of the simulations in Fig.\ref{fig:Loop_TOM} and Fig.\ref{fig:Loop_TOM_Diamond} that must be followed in order to calculate the TO and ThE terms:
\begin{itemize}[label=•]
  \item TO $\left(\frac{\partial \omega}{\partial  T}\right)_\alpha$: (a) $\rightarrow$ (b) $\rightarrow$ (d)   
  \item ThE $\frac{d T}{d \alpha}$: (a) $\rightarrow$ (b) $\rightarrow$ (c)
\end{itemize}

Notice that for the TO effect, we only need to measure the frequency shift due to temperature changes without any mechanical vibration or deformation. For the case of the ThE contribution, we want to quantify the material deformations caused by the thermal gradient field.  

\section{Numerical results}
\begin{figure}
    \centering
    \includegraphics[width=\columnwidth]{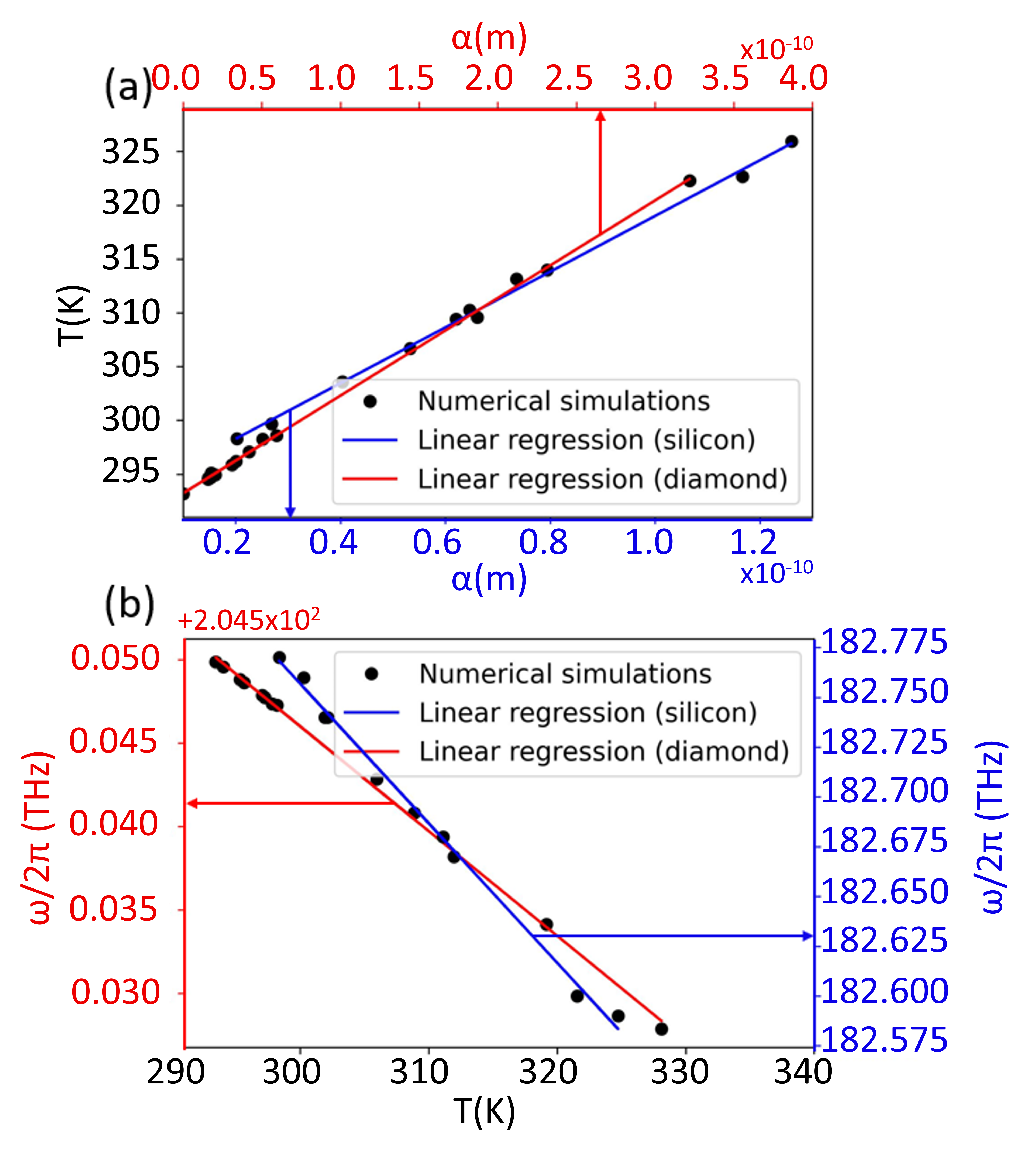}
    \caption{Data extracted from thermal simulations of the OM cavities showing the linear behavior of ThE (a) and TO (b) effects. (a) Maximum deformation with temperature of the cavity; (b) optical frequency shift with temperature without deformation. The data have been calculated by following the simulations in Figs.\ref{fig:Loop_TOM} and \ref{fig:Loop_TOM_Diamond}. The steps to obtain both the ThE and TO terms are described in Section IV.}
    \label{wttalpha}
\end{figure}

Here, we test the model by calculating and comparing the thermal contribution $g_0^{Th}$ to $g_0^{OM}$ in the two OM cavities introduced above with the parameters given in Figs.\ref{fig:allmodes} and \ref{fig:Montaje_2}.
Once we follow the paths described in the previous section to evaluate the TO and ThE effects, we only need to make a sweep increasing the power $Q_0$ that is absorbed into the cavity. That provides us with several points in the graphs $\omega-T$ and $T-\alpha$ respectively for the silicon and diamond OM cavities, with $T$ the maximum temperature value in the cavity. These results are shown in Fig.\ref{wttalpha}.  

After drawing all the points with their regressions, we can numerically calculate $\left(\frac{\partial \omega}{\partial  T}\right)_\alpha$ and $\frac{d T}{d \alpha}$ by taking the derivative of the curve that best fits the behavior of the points. One might notice that all tendencies are linear, and that implies constant values in all derivatives. Remarkably, this result is almost independent of the mechanical mode being analyzed, as the TO and ThE effects primarily depend on the material properties rather than the optical and mechanical modes. Hence, the results in Fig.\ref{wttalpha} are common for all mechanical modes shown in Figs.\ref{fig:allmodes} (b) and (c).
\begin{table*}
\centering
\begin{tabular}{|c|c|c|c|c|c|c|c|c|c|}
\hline & $X_{z p f}(m)$ & $\lambda(nm)$ & $\Omega / 2 \pi$ & $g^{P E} / 2 \pi(k H z)$ & $g^{M B} / 2 \pi(k H z)$ & $g^{O M} / 2 \pi(k H z)$ & $g^{T h} / 2 \pi(\mathrm{kHz})$ & $g_0^{OM} / 2 \pi(k H z)$ & $\begin{array}{l}g_0^{T h} / 2 \pi 
(\mathrm{kHz})\end{array}$ \\
\hline Silicon & $2.69 \times 10^{-15}$ & 1570 & $4.01 \mathrm{GHz}$ & -79797 & -4783 & -84580 & -4879 & -420 & -24 \\
\hline & $4.34 \times 10^{-14}$ & 1570 & $15.36 \mathrm{MHz}$ & -82 & 83150 & 83068 & $\begin{array}{l}-7872 \\
\end{array}$ & 413 & -39 \\
\hline Diamond & $6.74 \times 10^{-16}$ & 1470 & $5.95 \mathrm{GHz}$ & 14874 & 12462 & 27336 & -38 & 136 & -0.19 \\
\hline
\end{tabular}
\caption{Thermal contributions to the OM couplings of the silicon and diamond OM cavities under study.}
\label{tableg}
\end{table*}

\begin{figure}
    \centering
    \includegraphics[width=\columnwidth]{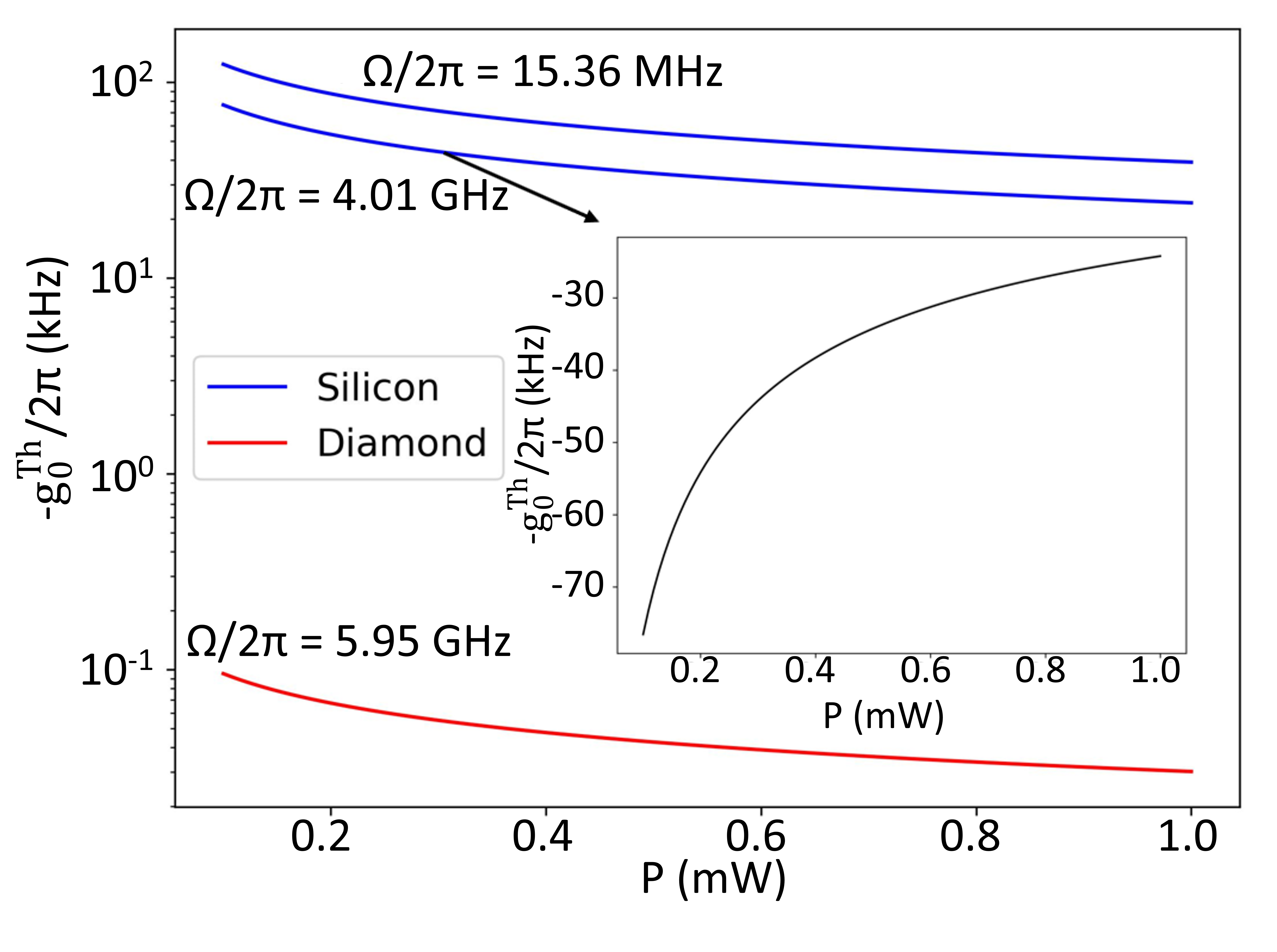}
    \caption{Thermal vacuum OM coupling rate $g_0^{Th}$ calculated using the simulated value of $g^{Th}$ for all mechanical modes of the silicon and diamond cavities. The figure is represented on a logarithmic scale to display all curves, with a negative sign applied as necessary. Inset: the $g_0^{Th}$ curve at $\Omega/2\pi=4.01\;GHz$ is included to illustrate the actual behavior of $g_0^{Th}$.}
    \label{fig:g0sim}
\end{figure}

From Fig.\ref{wttalpha}(a) we can obtain the ThE contributions, i.e. the slopes of the lines: $\frac{d T}{d \alpha}=2.59\times10^{11}\;K/m$ and $\frac{d T}{d \alpha}=9.07\times10^{10}\;K/m$ for the silicon and diamond cavities respectively. On the other hand, the TO contributions can be obtained from Fig.\ref{wttalpha}(b): $\left(\frac{\partial \omega}{\partial  T}\right)_\alpha=-4.40\times10^{10}\;Hz/K$ for the silicon cavity and $\left(\frac{\partial \omega}{\partial  T}\right)_\alpha=-3.95\times10^{9}\;Hz/K$ for the diamond cavity.\\
\indent 
Furthermore, we can compare the TO derivatives with a very simple calculation that relates the magnitude with the experimental TO coefficient\cite{Raghunathan:10}: $\frac{\partial \omega}{\partial  T}=-\frac{\omega}{n}k$, with $n$ the refractive index of the material and $k=\frac{\partial n}{\partial T}$ the TO coefficient. Therefore, by taking from literature\cite{MingkangNat,Corte2000,Raghunathan:10} the TO coefficient of silicon and diamond we obtain the results: $\frac{\partial \omega}{\partial  T}^{Formula}=-5.93\times10^{10}\;Hz/K$ for silicon and $\frac{\partial \omega}{\partial  T}^{Formula}=-8.33\times10^{9}\;Hz/K$ for diamond. Both results have the same order of magnitude. The differences with simulations are probably because the aforementioned formula is an approximation that considers changes in refractive index by the same proportion throughout the cavity. However, since we have an optical pattern, that assumption is not strictly correct. Nevertheless, our approach works nicely as a toy model and successfully predicts the order of magnitude of the effect in the cavities under study. We can then conclude that we can obtain a good approximation of the thermal contribution to the OM coupling rate from all these previous calculations.\\
\indent 
Indeed, considering the $X_{zpf}$ value for both cavities and using Eq.\ref{eq:g}, we can calculate $g^{Th}$, as shown in Table \ref{tableg}. We can see that we obtain a constant value of the thermal coupling rate $g^{Th}$ for each mechanical mode since all derivatives are also constant. This means that the thermal contribution correction is constant and does not depend on either the temperature or the optical power injected into the cavity. For that reason, in order to compare the thermal contribution with the combined PE and MB effects, the latter has been multiplied by a reasonable number of photons ($\bar{n}_{cav}=40400$). The reversed process has been applied to $g^{Th}$ in order to obtain $g_0^{Th}$ and compare it to $g_0^{OM}$, i.e. it has been normalized to $\bar{n}_{cav}=40400$, Eq.\ref{eq:g0}. That number corresponds to $1\;mW$ of absorbed power in the silicon cavity, and it has been calculated using Eq.\ref{fig:ncav} with the parameters: $Q=5915$, $\beta=0.26$ and $\lambda=1570\; nm$. As we can see in Table \ref{tableg}, $g^{Th}/2\pi$ is considerably smaller compared to $g^{OM}/2\pi$ for all mechanical modes.\\
\indent 
Furthermore, the comparison between the vacuum OM coupling rates $g_0^{OM}$ of all the mechanical modes with the vacuum thermal contribution $g_0^{Th}$ is represented in the graph $g_0^{Th}$-$P$, Fig.\ref{fig:g0sim}. In contrast to $g^{Th}$, we obtained a power-dependent value for the thermal vacuum OM coupling rate in Fig.\ref{fig:g0sim}, which means that at high power, the thermal correction to the OM coupling rate is lower than at low temperatures, since the correction is negative. The underlying mechanism of the power dependence in the OM coupling rate is related to the fact that higher optical powers lead to an increase in temperature, causing a change in the mechanical properties. Consequently, the cavity is no longer identical to its initial state, thereby influencing the coupling rate. This means that the total vacuum OM coupling rate $g_0^{Th}$ depends on the power injected, in agreement with the observations in \cite{Jiang:19}. We can also notice in Fig.\ref{fig:g0sim} that the trend is exactly the constant value $g^{Th}$ modulated by a square root function $\sqrt{\bar{n}_{cav}}$ for all the cases (see inset). Although this function has a maximum (negative) value for $\bar{n}_{cav}=1$, it is not representative since our model is classical, and it does not work for small numbers of photons where a quantum correction must be considered. In addition, in both silicon and diamond cavities, a minimum optical power in the cavity is required to transduce the optical and mechanical modes and then obtain the photothermal backaction. Therefore, our theoretical model makes sense for optical power at least of the order of hundreds of $\mu W$. It is also remarkable that for the GHz mode of the diamond cavity, the thermal correction is almost negligible ($0.13\%$ correction), whereas it becomes a $5.7\%$ correction for the GHz mechanical mode in silicon. This is consistent with silicon having much higher thermal effects than diamond.  

\section{Conclusions}
In this work, we have presented a theoretical approach to studying the OM coupling in OM crystal cavities that takes into account the influence of the thermo-optic and thermal expansion effects. By considering these thermal contributions, we have extended the conventional OM coupling equation, which traditionally includes only the photoelastic (PE) and moving boundaries (MB) effects, to a more comprehensive model that encompasses the complete dynamical backaction loops arising from the optomechanical and photothermal effects. Using numerical simulations in \textit{Comsol Multiphysics}, we have obtained the TO and ThE contributions for silicon and diamond OM crystal cavities. We have found that the TO and ThE effects exhibit a linear behavior, and their values are almost independent of the mechanical modes being analyzed, primarily depending on the material properties. From these simulations, we calculated the thermal OM coupling rate $g^{Th}$, which represents a thermal correction to the conventional OM coupling rate. Comparing $g^{Th}$ with the conventional OM coupling rates ($g^{PE}$ and $g^{MB}$) multiplied by a reasonable number of intracavity photons, we found that the thermal contribution is significantly smaller than the conventional contributions for all mechanical modes.

Furthermore, we have analyzed the thermal vacuum OM coupling rate $g_0^{Th}$, which depends on the power injected into the cavity. We observed that $g_0^{Th}$ follows a square root dependence on the average number of photons in the cavity, $\bar{n}_{cav}$, for all mechanical modes. The fundamental reason behind the power dependence in the OM coupling rate is that higher optical powers raise the material temperature, therefore altering mechanical properties and deforming the cavity, ultimately impacting the OM coupling rate. However, this behavior is only valid for powers starting from hundreds of $\mu \text{W}$, since quantum corrections should be considered for lower photon numbers. Although we have studied thermal effects on silicon and diamond cavities, the same approach could be followed for cavities made of other materials such as gallium arsenide or lithium niobate\cite{MBPE,Jiang:19,McCaulley1994}.

Moving forward, an important future direction would be to experimentally measure this thermal correction, which will require higher precision in measuring the OM coupling rate to ensure that the measurement error does not exceed the thermal contribution magnitude. By precisely characterizing the thermal contributions to the OM coupling rates, researchers could better understand and control the thermal effects in optical microcavities. This experimental validation would also enable the refinement and optimization of OM devices for a wide range of applications, including sensing \cite{GilSantos2020}, signal processing \cite{NAV17_NCOMMS}, and quantum applications such as quantum information processing\cite{Mirhosseini2020,PhysRevLett.109.013603} or quantum nonlinearities \cite{QN1,QN2}.

\section*{Data Availability Statement}

The data that support the findings of this study are available from the corresponding author upon reasonable request.

\begin{acknowledgments}
The authors thank Jeremie Maire, Nestor Capuj, Carles Milian, and Ewold Verhagen for useful discussions. The authors acknowledge funding from "Generalitat Valenciana" (GVA, CIACIF/2021/006 ), the European Union (“NextGenerationEU”/PRTR and “ERDF A way of making Europe”), and the Spanish Ministry of Science and Innovation (MCIN/AEI/10.13039/501100011033) under project grants PID2021-124618NB-C21 (ALLEGRO) and PCI2022-135003-2 (MUSICIAN).
\end{acknowledgments}

\section*{References}

%

\end{document}